\documentstyle[11pt]{article}
\newtheorem{definition}{Definition}[section]

\newtheorem{theorem}[definition]{Theorem}

\newcommand{\comment}[1]{}

\begin{document}

\title{The Computational Complexity\\ of Generating Random Fractals}

\author{Jonathan Machta
\\
Department of Physics and Astronomy\\
University of Massachusetts\\
Amherst, Massachusetts 01003\\
e-mail address: machta@phast.umass.edu
\and
Raymond Greenlaw
\\
Department of Computer Science \\
University of New Hampshire \\
Durham, New Hampshire 03824\\
e-mail address: greenlaw@cs.unh.edu\\[.5in]
}

\maketitle

\begin{abstract}
In this paper we examine a number of models that generate random
fractals.  The models are studied using the tools of computational
complexity theory from the perspective of parallel computation.
Diffusion limited aggregation and several widely used algorithms for
equilibrating the Ising model are shown to be highly sequential; it is
unlikely they can be simulated efficiently in parallel. This is in
contrast to Mandelbrot percolation that can be simulated in constant
parallel time.  Our research helps shed light on the intrinsic
complexity of these models relative to each other and to different
growth processes that have been recently studied using complexity
theory.  In addition, the results may serve as a guide to simulation
physics.

\end{abstract}

\section{Introduction}

Random fractals are a major focus of investigation in statistical
physics.  Such patterns occur at equilibrium critical points and arise
through a variety of non-equilibrium dynamical processes.  A number of
models generate random fractals including diffusion limited
aggregation (DLA) and the Ising model at criticality. These models
have been extensively studied by computer simulation methods and, in
some sense, they are defined by the algorithms that are used to
simulate them.  In this paper we examine such defining algorithms from
the viewpoint of the theory of computational complexity.

{\em Computational complexity\/} is the branch of theoretical computer
science that seeks to quantify the resources required to solve
problems.  One of the main achievements of complexity theory is the
identification of a hierarchy of complexity classes.  The classes
differ with respect to how the various resources, such as time, space
and processors, scale in proportion to problem size.  For example,
does the running time increase as a logarithmic, power or exponential
function of the problem size?  Our emphasis is on parallel
computational complexity.  We seek to answer the following question:
how do the number of processors and the amount of time required to
simulate a system on a massively parallel computer increase with the
system size?

The motivation for this work is two-fold.  First, computational
complexity may serve as a guide to simulation physics.  With the
growing availability of massively parallel computers, it is important
to investigate models from the perspective of parallel complexity.
Another, perhaps more significant, motivation is to provide an
alternative characterization of these models.  An enormous amount of
effort has gone into characterizing the morphology of fractal patterns
via critical exponents, fractal and multifractal dimensions, scaling
functions, and so on.  Such characterizations fail to adequately
distinguish these models from the standpoint of what can be described
intuitively as complexity.  We believe that the intuitive notion of
physical complexity is at least partially captured by the
computational complexity measure of parallel time (with the number of
processors appropriately restricted).  This idea, in a slightly
different form, has been previously proposed by Bennett~\cite{Benn90}.

In a nutshell, the idea is that simple objects can be generated
quickly while complex objects require a long history for their
formation.  We illustrate this by comparing two random fractals.  The
first is Mandelbrot percolation~\cite{Man83}; an example of which is
depicted in Fig.~\ref{manperc}.  We show Mandelbrot patterns require
only (non-uniform) constant parallel time to generate. Though they are
fractals, there is very little interesting morphology; the structure
on each length scale is independent of the structure on other length
scales.  Many properties of Mandelbrot percolation are susceptible to
rigorous analysis \cite{ChChDu}.  The second example is
DLA~\cite{WiSa} that generates fractal patterns like those shown in
Fig.~\ref{dlareal}.  DLA patterns are produced by a highly sequential
algorithm that seems to require polynomial (in the size of the
aggregate) parallel time.  DLA patterns reflect a subtle interplay of
randomness and structure on many length scales.  DLA has remained
largely refractory to theoretical analysis.  Whether or not one
accepts a definition of physical complexity in terms of computational
complexity, it is interesting that a variety of models in statistical
physics can be sharply separated from one another by a fundamental new
yardstick.

Our research extends a study of the complexity of a number of growth
models.  Ref.~\cite{Mac93a} is concerned with a fluid invasion model
that generates clusters with the same statistics as DLA\@. This model
is shown to be inherently sequential (technically, {\bf P}-complete)
and so it is unlikely that it can be efficiently simulated in
parallel.  Here we show that the original random walk dynamics for
generating DLA clusters is also inherently sequential.  In
Ref.~\cite{MaGr} we considered a number of other growth
models---invasion percolation, Eden growth, ballistic deposition and
solid-on-solid growth---and showed that all of these models can be
efficiently simulated in parallel.  The fractal patterns associated
with them can be generated on a parallel computer in a time that
scales logarithmically in the system size while using a reasonable
number of processors.  Although each of these models is less complex
than DLA, each is more complex than Mandelbrot percolation.

Other applications of computational complexity theory to statistical
physics have focused mainly on the existence of polynomial time
sequential algorithms.  For example, the problems of finding the exact
ground states of spin glasses~\cite{Bara} and computing self-avoiding
walks in a random environments~\cite{Mac92a} have been shown to be
computationally intractable (technically, {\bf NP}-complete). On the
other hand, a polynomial time algorithm exists for the random field
Ising model~\cite{AnPrRa}. There is also work that establishes the
complexity of finding the partition function of the Ising model and
related spin models on arbitrary lattices either exactly~\cite{Welsh}
or approximately using Monte Carlo methods~\cite{JeSi}.

In Section~\ref{se-complexity} we give an introduction to
computational complexity theory.  A reader familiar with this field
may want to skim this section.  In Section~\ref{se-fractals} we
investigate the computational complexity of the following systems:
Mandelbrot percolation, DLA, Metropolis dynamics for the Ising model,
Wolff dynamics for the Ising model, and Swendsen-Wang dynamics for the
Ising model.  Section~\ref{se-discussion} is devoted to a discussion
of the results.

\section{Computational Complexity Background}
\label{se-complexity}

In this section we provide an introduction to computational
complexity theory.  The reader can find further information and
details in a number of texts~\cite{hu79,lp81} and
monographs~\cite{GrHoRu,Jo90,GaJo}.

\subsection{The parallel random access machine}

The theoretical model we focus on is the parallel random access
machine or P-RAM\@.  It is the most commonly used model in parallel
computation.  We describe the P-RAM and then relate its resource usage
to the corresponding measures for actual parallel computers.

The P-RAM consists of a number of processors each with local memory
and having access to a common global random access memory.  All
processors run the same program but are distinguished by non-negative
integer labels so that the processors may operate on their own data or
skip instructions.  Input to the machine is placed in designated,
consecutive global memory locations as is output. The P-RAM is in the
class of single-instruction multiple-data-stream (SIMD) models.  The
processors run synchronously and in each time step a single {\em
random access machine\/} (RAM) instruction~\cite{AhHoUl} or a global
memory access instruction is executed by a subset of the processors.
Examples of typical instructions are `write the contents of the
accumulator to memory location $a$' and `add the contents of the
accumulator to the contents of register $a$, placing the sum in the
accumulator.'

Although many processors may read the same memory location at a
particular time, difficulties arise if multiple processors attempt to
write to the same location.  One frequently used arbitration scheme is
the concurrent write model in which processors are assigned a write
priority.  When more than one processor attempts to write to a given
location, the processor with the highest priority succeeds.  This
model is known as the PRIORITY CRCW P-RAM~\cite{Fi}.  We adopt this
model and simply refer to it as the P-RAM\@.

In the P-RAM model any processor can access any global memory
location in one time step; the model allows unlimited parallelism.
For this reason the P-RAM serves as a convenient model for designing
and analyzing parallel algorithms, for studying processor and time
requirements and for proving lower bounds. Although the P-RAM is
overly simplistic in its assumptions, it can nevertheless be
simulated on models of parallel computation with more restricted
connectivity such as the hypercube. These simulations usually have a
slow down of a logarithmic factor and require roughly the same
amount of hardware as the corresponding P-RAM computations,
see~\cite{KaRa} for additional details and references.

As an example of the utility of parallelism, consider the task of
computing the {\em parity\/} of $n$ bits.  Parity is the problem of
determining whether there is an even number of 1's in the input.
Initially, the $n$ bits are stored in global memory locations $1,
\ldots, n$.  A P-RAM program that computes parity uses $n/2$
processors numbered $1, \ldots, n/2$ to add $n/2$ pairs of bits
(modulo 2) in parallel.  That is, processor~1 adds the contents of
locations 1 and 2 storing the result (modulo 2) in location~1,
processor~2 adds locations 3 and 4 storing the result (modulo 2) in
location~2, and so on.  Similarly, the resulting $n/2$ values are
added pairwise (modulo 2) by $n/4$ processors.  This process is
repeated until, after $\lceil \log_2 n
\rceil$ time steps, parity is computed.  Notice, in this case the
algorithm's output is placed in global memory cell 1 by processor
number 1.  The algorithm runs in $O(\log n)$ time using $n/2$
processors.

\subsection{Complexity classes}

The primary question addressed by computational complexity theory is
how the difficulty of a computation scales with the size of the
problem instance.  Complexity theory usually focuses on {\em decision
problems}.  An instance of a decision problem is a string of bits
encoding the problem; the solution is simply a 1 or 0.  If the
solution for input $x$ is 1, we say that $x$ is `accepted' and
otherwise $x$ is `rejected.'  In this sense, a decision problem is
defined by its set of accepted strings.  The problem size, $n=|x|$, is
the length of the encoded input.  A simple example involving the
parity problem discussed above is as follows:\\

\noindent
{\bf Parity}\\
{\bf Given:} $b_1, \ldots, b_n$, where $b_i \in \{ 0, 1\}$.\\
{\bf Problem:} Do an even number of the $b_i$'s have value 1?\\

\noindent
In this case the input is easily encoded using exactly $n$ bits.  The
output is a 1 if there are an even number of $b_i$'s with value 1 and
0 otherwise.  It is easy to see that the answer may be found on a
single processor computer (such as a RAM or more familiar desktop
computer) with a running time that scales linearly in $n$ by simply
scanning through the bits and maintaining their sum modulo 2.

We now define several important complexity classes for parallel
computation.

\begin{definition} \hspace*{1in}\\[-.2in]
\label{def-classes}
\begin{itemize}
\item
The class {\bf AC}$^0$ consists of those decision problems that can be
solved on a P-RAM in $O(1)$ (constant) time using $n^{O(1)}$
(polynomial) processors.
\item
The class {\bf NC} consists of those decision problems that can be
solved on a P-RAM in $(\log n)^{O(1)}$ (poly-logarithmic) time using
$n^{O(1)}$ processors.
\item
The class {\bf P} (polynomial time) consists of those decision
problems that can be solved on a P-RAM in $n^{O(1)}$ time using
$n^{O(1)}$ processors.
\end{itemize}
\end{definition}

\noindent
It is easy to see that {\bf AC}$^0$ $\subseteq$ {\bf NC} $\subseteq$
{\bf P}\@.  It is known that {\bf AC}$^0$~$\neq$~{\bf NC} and, while
no proof yet exists, it is widely believed that {\bf NC}~$\neq$~{\bf
P}\@.  The classes in Def.~\ref{def-classes} are robust in the
sense that they may be equivalently defined with respect to several
different computation models~\cite{Jo90}; they are not tied to the
P-RAM model of parallel computation.

All of the problems considered in this paper are in the class {\bf
P}\@.  It is generally accepted that problems in the class {\bf P}
have feasible sequential time solutions.  The question we pose for the
fractal models is whether a polynomial time problem can be
qualitatively sped up via massive parallelism.  For the parity
example, the sequential solution mentioned takes $O(n)$ time.  The
parallel solution outlined previously shows a P-RAM can solve this
problem in $O(\log n)$ time using $n/2$ processors.  Thus, the parity
problem is in the class {\bf NC} and a qualitative speed-up is
achieved in the parallel setting.  On the other hand, parity is {\em
not\/} in {\bf AC}$^0$, see~\cite{Fi} for example.

We will use the terminology that problems in {\bf NC} (and thus {\bf
AC}$^0$) are `efficiently solved in parallel,' since we obtain a
qualitative speed-up solving these problems in parallel.  On the other
hand, problems that are in {\bf P} but likely not in {\bf NC} are
called `inherently sequential.'  The running time for solving an
inherently sequential problem cannot be decreased from polynomial to
poly-logarithmic using a polynomial number of processors.  Below we
identify a class of problems that are in {\bf P} but not {\bf NC}
(unless it happens that {\bf NC}~$=$~{\bf P})\@.

In order to proceed we need to be able to relate problems to one
another.  This is accomplished via the notion of reduction.  The idea
is similar to a commonly used programming practice.  To solve one
problem, we often use a subroutine call to a different problem.  In
this sense we reduce our original problem to the one involved in the
subroutine call.  More formally,

\begin{definition}
Let $n = |x|$.  (Throughout, $|x|$ denotes the length of string $x$
and a decision problem $D$ is represented as a set of accepted
strings.)  Decision problem $D_1$ is {\bf NC many-one reducible} or
{\bf NC reducible} ($\prec$) to decision problem $D_2$ if there exists
a function $f$ such that $x
\in D_1$ if and only if $f(x) \in D_2$, and $f$ can be computed on a
P-RAM in $(\log n)^{O(1)}$ time using $n^{O(1)}$ processors.
\end{definition}

\noindent
If $D_1\prec D_2$ then $D_1$ is `no harder' than $D_2$\@.  This is
because we could solve $D_1$ using an algorithm for $D_2$, where the
input to $D_2$ is produced by an efficient calculation involving
$f$. We can also compare a given problem to an entire complexity
class, via the concept of `completeness.'

\begin{definition}
A decision problem $D$ is {\bf P}-{\bf complete} if \\
\hspace*{1.3in} (1) $D \in {\bf P}$ and\\
\hspace*{1.3in} (2) for all $D' \in {\bf P}$, $D' \prec D$.
\end{definition}

The {\bf P}-complete problems are therefore the hardest problems in
{\bf P}\@.  Based on these definitions, the following theorem is
straightforward to obtain.

\begin{theorem}
If any {\bf P}-complete problem is in {\bf NC} then {\bf
NC}~$=$~{\bf P}.
\end{theorem}

\noindent
Thus, if the well-known conjecture in computer science that {\bf
NC}~$\neq$~{\bf P} holds (and there is lots of evidence supporting
this conjecture~\cite{GrHoRu}), {\bf P}-complete problems are {\em
inherently sequential}.

$\prec$ is a transitive relation.  That is, if $D_1 \prec D_2$ and
$D_2 \prec D_3$, then $D_1 \prec D_3$.  Therefore, if some problem
$D'$ is shown to be {\bf P}-complete and $D' \prec D$ then $D$ must be
as difficult to solve as $D'$\@.  In this case we say $D$ is {\bf
P}-{\em hard}.  If $D$ is also in {\bf P}, it is {\bf P}-complete as
well.  Using transitive reductions, a large number of {\bf P}-complete
problems have been identified and no efficient parallel solution has
been found for any of them, providing evidence for the conjecture.  In
this paper we will prove that several problems from statistical
physics are {\bf P}-complete by showing that known {\bf P}-complete
problems reduce to them.

The fundamental {\bf P}-complete problem is the circuit value problem
(CVP); it is phrased in terms of Boolean circuits.  Before describing
CVP, we give an informal description of circuits.  A Boolean circuit
is a collection of connected {\sc not}, {\sc and} and {\sc or} gates.
{\sc not} gates have one input and multiple outputs; {\sc and} and
{\sc or} gates have multiple inputs and multiple outputs.  The {\em
fan-in\/} ({\em fan-out\/}) is the number of inputs (outputs) of a
gate.  The connection of the gates is `feedforward.'  That is, it must
be possible to number the gates so that the outputs of a gate are
connected to the inputs of gates with higher numbers.  Such a
numbering is called a {\em topological numbering\/} and we say the
gates are in topological order.  In calculating outputs from inputs
each gate computes its Boolean function just once.  Sometimes gates
other than {\sc not}, {\sc and} and {\sc or} are considered.  The {\em
size\/} of a circuit is defined as the number of gates.  The {\em
depth\/} is the longest path from an input to an output.\\

\noindent
{\bf Circuit value problem} (CVP)\\
{\bf Given:} A compact\footnote{A {\em compact encoding\/} of a circuit is
polynomial in the circuit size.} encoding $\overline{\alpha}$ of a
Boolean circuit together with its inputs $x_1, \ldots, x_n$, and a
designated output gate $g$.\\ {\bf Problem:} Does $g$ evaluate to 1
on input $x_1, \ldots, x_n$?

\begin{theorem}
The circuit value problem is {\bf P}-complete~\cite{lad75}.
\end{theorem}

Numerous variants of CVP are {\bf P}-complete~\cite{GrHoRu}.  In {\sc
nor} CVP the circuit consists entirely of {\sc nor} gates with fan-in
and fan-out two.  {\sc nor} CVP without fan-out restrictions is also
{\bf P}-complete for planar circuits; this version is called planar
{\sc nor} CVP\@.  In monotone CVP the circuit is composed of {\sc and}
and {\sc or} gates ({\sc not} gates are absent) having fan-in and
fan-out two. This problem is {\bf P}-complete for arbitrary circuits
but it is in {\bf NC} for planar circuits.  We shall make use of
planar {\sc nor}, monotone and other restricted variants of CVP in
Section~\ref{se-fractals}.

\comment{Anderson's thesis, page 6 shows OR, NOT layered CVP
is P-complete.  I don't think the planar, NOR, layered CVP is
P-complete. 1/10/95 RG}

A proof that a problem $D$ is {\bf P}-complete via a reduction from
CVP is tantamount to what in other contexts has been called
`computational universality.'  The dynamics of hard spheres in
classical mechanics~\cite{FrTo} and some cellular automata
rules~\cite{GrHoRu,VoBu} have been shown to be computationally
universal.  Our proofs that DLA and various Ising Monte Carlo dynamics
are {\bf P}-complete depend on showing that arbitrary logical
calculations can be embedded in these dynamics.

In addition to the resources of parallel time and number of
processors, the notion of {\em uniformity\/} plays an important role
in computations.  Roughly speaking, a uniform solution to a problem
uses the ``same'' program for each problem size whereas a non-uniform
solution may use a different program for each size.  For example,
simulating the Ising critical point in three dimensions using
conventional Monte Carlo methods is a non-uniform problem because the
critical temperature is required as a parameter in the algorithm.  As
the system size increases, the program must contain an increasingly
accurate value of the critical temperature.\footnote{Specifically, the
critical temperature must be known to accuracy $L^{-1/\nu}$, where
$L$ is the system size and $\nu$ is the correlation length
exponent.}  On the other hand, simulating DLA clusters is a uniform
task since no fine tuning of parameters is required.  The same can
be said for other `self-organized' critical points such as invasion
percolation.  Recently, a uniform algorithm for sampling Ising
critical points has been developed~\cite{MaChCa}.

\subsection{Parallel time and logical depth}

The P-RAM model and the complexity measures that are built from it are
in some sense unphysical because unit time is assigned to a single
read or write step.  Eventually, as such a device is scaled up, the
communication time between processors and memory dominates the running
time and the unit time assumption fails. Indeed, in the limit of large
systems all of the models discussed here require polynomial time to
simulate on any real world device because all are capable of
generating random patterns with correlations on the scale of the
system size.  These correlations cannot be set up without
communication across the system and this requires polynomial time.
\comment{Here we are taking into account wire length.  RG 1/6/95
Yes, JM 1/18/95}

Nonetheless, parallel time correctly identifies an important aspect of
the problem which can be called `logical depth.'  The logical depth is
the minimum number of logical operations that must be carried out in
sequence before a problem is solved.  This concept can be made
rigorous by considering families of Boolean circuits~\cite{Jo90}.  A
family of Boolean circuits, one for each problem size, can simulate a
P-RAM programmed to solve a given problem and vice versa.  The
definitions of the complexity classes {\bf AC}$^0$, {\bf NC} and {\bf
P} can be stated in terms of families of Boolean circuits: the number
of processors corresponds roughly to the size of the circuit (number
of gates) and the parallel time roughly to the depth of the circuit
(length of the longest path from input to output).  Thus for example,
a problem is in the class {\bf NC} if it can be solved by a uniform
family of Boolean circuits having polynomial size and poly-logarithmic
depth in the number of inputs (the problem size).

A few comments are in order regarding the number of processors.  If
only one processor is allowed, then all the problems treated here
require polynomial time.  If on the other hand, the number of
processors is unrestricted it can be shown~\cite{Ve} that all the
problems discussed here are solvable in constant P-RAM time using
exponentially many processors (or equivalently by circuit families
with exponential size and constant depth) and again the interesting
distinctions based upon parallel time disappear.  Interesting results
are found when polynomial parallelism is permitted.

\subsection{Complexity of sampling methods}
\label{se-sampling}

Computer scientists study decision problems whereas computational
statistical physicists are usually concerned with sampling
problems---generating states from some equilibrium or nonequilibrium
distribution.  Sampling algorithms require a supply of random numbers
and produce as output a system configuration.  This configuration is
described by $m$ bits representing the degrees of freedom of the
system expressed in binary.  One can extend the ideas of complexity
theory to sampling methods by introducing probabilistic P-RAM's in
which each processor is equipped with a register for generating random
bits.

Instead of producing random bits dynamically one could equivalently
produce the required random bits in advance and include them as inputs
to a deterministic calculation.  In this way a sampling method is
reduced to $m$ decision problems, one for each binary degree of
freedom.  An example of such a decision problem is `Does Ising spin
$s_j$ ($1 \leq j \leq m$) have value $+1$ after $M$ iterations of the
Monte Carlo procedure using random numbers $x_i$?'
Note that these $m$ decision problems may be run in parallel with, in
the worst case, a factor of $m$ increase in the number of
processors. {\em Therefore, the sampling algorithm has the same
parallel time requirement up to a constant factor as the associated
decision problem.}

In statistical physics, the problem size is conventionally identified
with the system size; the number of bits, $m$ required to specify a
system configuration.  This differs from complexity theory where it is
the number of bits required to state the problem that is identified as
the problem size.  The following definition insures that the two
notions of problem size are compatible.  For a given sampling method
with $r$ random inputs, $o$ ordinary inputs and $m$ outputs, we define
the associated natural decision problem as follows.  The input is of
length $m+o+r$.  The first $m$ bits represent the degrees of freedom
of the system.  Of these bits exactly one is a 1.  The position of the
1 specifies which degree of freedom of the system (e.g. which Ising
spin) is to be evaluated.  Since the selected degree of freedom is
expressed in {\em unary}, the decision problem size is at least as
great as the system size.\footnote{This helps insure that the problems
considered are in {\bf P} and that the number of processors used will
be polynomial in the input size.}  For example, to represent the fifth
out of ten degrees of freedom our unary expression would be
`0000100000.'  The next $o$ bits are the ordinary inputs to the
problem expressed in a suitably compact form.  These inputs might
include the size of the lattice, the temperature, the number of
iterations of an elementary Monte Carlo step and other relevant
parameters expressed in binary notation.  The final $r$ bits are the
random bits needed for the sampling method.  So that the answer or
other potentially useful information is not built into these bits, we
require that they be interpreted as independent random variables that
take the value 1 with probability $1/2$.  We restrict our attention to
`reasonable' sampling methods where $r$ is bounded by a polynomial in
$m$.

The decision problem for a sampling method can now be studied using
conventional computational complexity theory.  It must be emphasized
that the complexity of the decision problem is only an upper bound on
the complexity of sampling a given distribution.  The reason is that
the decision problem is associated with a particular sampling
method. It may be that an alternative method leads to a less complex
decision problem.  In principle we would like to know how the time,
number of processors and number of random bits scale with $m$ for the
optimal sampling method.  Unfortunately, tools for studying this
question have yet to be developed.  Instead, we focus on the
complexity of several known sampling methods. Nonetheless, if the best
known sampling methods are investigated and their complexity is
established, it is plausible that the complexity of sampling has also
been found. (Note, proving that a particular sampling method is
optimal seems to be a very difficult task.)

\section{Complexity of Random Fractals}
\label{se-fractals}

In this section we consider the following models: Mandelbrot
percolation, diffusion limited aggregation and the Ising model.  We
discuss sampling methods for these systems and then study the parallel
computational complexity of the associated decision problems.  Each
model generates random {\em mass fractals}---sets of `occupied' sites
whose number scales as a noninteger power of the lattice size.

\subsection{Mandelbrot percolation}
\label{se-mp}

This random fractal was first described by Mandelbrot~\cite{Man83}.
It was analyzed by rigorous methods in~\cite{ChChDu}, and was later
generalized and applied as a model of a fractal porous
medium~\cite{Mac91a,ChChMa,ChMa}.  Mandelbrot percolation is defined
on a $d$-dimensional lattice.  It is parameterized by a rational
retention factor $Q$ ($0 \leq Q<1$), a positive integer rescaling
factor $N$ and iteration number $k$.  System
configurations are described by a bit at each lattice site.  If the
bit is a 1, we say the site is `occupied.' For purposes of
illustration, we consider the two dimensional version on an $N^k
\times N^k$ square lattice.  A configuration is generated in the
following way: at the $i^{\rm th}$ step ($0 \leq i
\leq k-1$) the lattice is completely divided into, $N^i \times N^i$
non-overlapping squares and each square is independently `retained'
with probability $Q$\@.  If a square is retained, the site(s) in it
are not changed.  If a square is not retained then all of the site(s)
in it are changed to unoccupied.  After $k$ steps unoccupied regions
with a wide range of sizes are typically created. The resulting set of
occupied sites is a random fractal with limiting Hausdorf dimension,
$D_H=2+(\log Q)/(\log N)$ if $D_H >0$. A realization of Mandelbrot
percolation with $N$~$=$~$2$, $Q$~$=$~$.9$ and $k$~$=$~7 is shown in
Fig.~\ref{manperc}.

A natural decision problem associated with Mandelbrot percolation
takes as input random numbers, $x_i$ with $0 \leq x_i<1$.  These
numbers are used to generate `retention bits' that are 1 if $x_i <
Q$ and 0 otherwise.  Each retention bits determines whether a
particular square of a given size is retained.\\

\noindent
{\bf Mandelbrot percolation} (dimension $d$, scale factor
$N$, precision $b$)\\
\noindent
{\bf Given:}  A non-negative integer $k$, a designated lattice site
$s$ expressed in unary with $|s| = N^{dk}$,
a retention factor $Q$ ($0 \leq Q < 1$) with $Q$ represented
by a $b$-bit binary number\footnote{Our method of producing
random numbers via coin tossing suggests this coding choice.
Such a scheme does not allow all possible rationals in the interval
$[0,1)$ to be represented.} and a list of $(N^{dk}-1)/(1-N^{-d})$
random numbers $x_i$ with $0 \leq x_i<1$ expressed as a $b$-bit
number.\\
\comment{Other possibilities include a more general random number
generator where we pick a number from a range instead of toss a
coin, or we are given bits that have a probability of Q of being
1 and 1-Q of being 0.  RG 1/28/95}
{\bf Problem:} Is site $s$ occupied by the Mandelbrot
percolation process?\\

The instances of Mandelbrot percolation require that the dimension,
scale factor and precision are all fixed inputs.  In terms of
the discussion of Section~\ref{se-sampling} relating decision and
sampling problems, $|s| = m$, $\lceil \log_2 k \rceil + b = o$ and $b
(N^{dk}-1)/(1-N^{-d}) = r$.\footnote{This is not precise as
delimiters are also used in the encoding to make decoding easier.}

A constant time P-RAM algorithm for Mandelbrot percolation is sketched
below.  First, retention bits for every square of each size are
computed in parallel by comparing the $x_i$ to $Q$\@.  Since $b$ is a
constant, this can be done in constant time.  The $m$ retention bits
for the individual sites are placed in memory cells 1 to $m$.  For
each site $j$ ($1 \leq j \leq m$) the occupancy of $j$ is determined
by taking the {\sc and} of all the retention bits of the $k$ squares
containing $j$.  To compute the {\sc and}, all processors reading a
retention bit 0 write a value of 0 into global memory cell $j$.  This
step uses $km$ processors.  Note, cell $j$ is 1 if site $j$ is
occupied and 0 otherwise.  Next, the {\sc and} of cell $j$ and the
$j^{\rm th}$ place in the unary expression of $s$ is computed; the
result placed in cell $j$.  Now, cell $j$ is 1 if and only if site $j$
is occupied and is the selected site.  Finally, the {\sc or} of cells
1 through $m$ is taken (by having any processor reading a 1 write to
memory cell 1) to determine if the selected site is occupied.

This P-RAM algorithm uses constant time and polynomial ($km$)
processors.  For problems in constant time a very strong notion of
uniformity (meaning highly uniform), DLOGTIME, is typically imposed.
The algorithm described does not seem to be DLOGTIME uniform.  This is
because for each lattice size, a different program may be needed to
quickly incorporate information about which retention bits belong to a
given site.  Note that in general proving a problem does not meet a
given uniformity condition is very difficult.  Our algorithm is {\bf
NC}$^2$ uniform.  We have the following:

\comment{I think this weakly non-uniform AC^0 is NC,  JM 1/23/95
I agree.  It looks like it may be in NC^2 if not NC^1.  RG 1/30/95}

\begin{theorem}
\label{th-mp}
Mandelbrot percolation is in non-uniform ${\bf AC}^0$.
\end{theorem}

\subsection{Diffusion limited aggregation}

Diffusion limited aggregation~\cite{WiSa} is a cluster growth model
where new occupied sites are added to the growing cluster one at a
time. Here we illustrate DLA for a two dimensional lattice with growth
initiated along a line.  A random walker is started at a random
position along the top edge of an $L \times L$ square lattice.  The
walker moves until it is a nearest neighbor of an existing occupied
site at which point it joins the cluster.  Initially, the bottom
edge of the lattice is considered occupied.  If a walk fails to join
the cluster, hits the top boundary of the lattice or is unable to
move (goes off the lattice or encounters a site that is occupied in
its first move), it is discarded. A new random walk is started as
soon as the previous walk has joined the cluster or been discarded;
the process continues until a cluster of the desired size is grown.

A natural decision problem associated with the dynamics of diffusion
limited aggregation is defined below.  \\

\noindent
{\bf Diffusion limited aggregation} (dimension $d$)\\
{\bf Given:} Three positive integers $L$, $M_1$ and $M_2$,
a designated site $s$ expressed in unary
with $|s| = L^{d}$ and a list of
random bits specifying $M_1$ walk trajectories each of length $M_2$
defined by a starting point on the top edge of the lattice together
with a list of directions of motion (e.g. N, S, E and W for two
dimensions).\\  {\bf Problem:} Is site $s$ occupied by the
aggregation process?\\

\noindent
The proof that DLA is {\bf P}-complete proceeds by a reduction from a
variant of the planar {\sc nor} circuit value problem.  The reduction
has a similar flavor to the proof that a closely related fluid
invasion problem is {\bf P}-complete~\cite{Mac93a}, although there
seems to be no way to make use of that proof directly.

\begin{theorem}
\label{th-dla}
Diffusion limited aggregation is {\bf P}-complete.
\end{theorem}

{\em Proof sketch:\/} The idea is to prescribe a sequence of walks
capable of carrying out the evaluation of a modified (but still {\bf
P}-complete) version of the planar {\sc nor} circuit value problem.
In this version of CVP the {\sc nor} gates have a fan-in and fan-out
of two.  We also allow single input {\sc or} gates with fan-out
restricted to at most two.  The circuit encoding requires that the
gates are numbered in topological order.  The encoding specifies a
planar layout of the circuit with gates being located at grid points.
Finally, the circuit is required to be {\em synchronous}.  That is,
each gate receives its inputs only from gates on the immediately
preceding level.  Gates at level one are the only gates that are
allowed to have direct circuit inputs.  It can be shown that this
version of CVP is {\bf P}-complete.

The walks to simulate the circuit are chosen so that the cluster grows
along linear paths of sites and bonds that play the role of wires
connecting gates.  A wire carries the value {\sc true} if the cluster
grows along it.  Wires that remain unoccupied carry the value {\sc
false}\@.  The gates themselves are represented by locations where
wires meet and several parts of the growing cluster interact.  Below
we describe how logical values are propagated along wires and how
{\sc nor} and {\sc or} gates are implemented.

Logical values are propagated as follows.  Each
wire is realized by a preassigned sequence of walks.  These walks move
to successive locations along the wire.  Each walk moves to its
assigned site along the wire and, if the value of the wire is {\sc
true}, it sticks there.  Each walk reverses its path after reaching
its assigned site.  In this way if the wire carries the value {\sc
false}, the walk returns to the upper boundary and is discarded. For
example, the first walk creating the output wire for the gate shown in
Fig.~\ref{dlanor} arrives at site `d' from above. If `c' is occupied
this walk sticks at `d' and the cluster begins to grow along the
output wire.  If `c' is not occupied, the walk turns around and
retraces its steps back to the upper boundary where it is discarded.
Thus the cluster grows along the output if site `c' is occupied.  It
is straightforward to have the output wire split into two separate
wires.  These then become the inputs to other gates.  Note, a larger
fan-out could be supported, however, the details of the proof become
more involved.

A {\sc nor} gadget is shown in Fig.~\ref{dlanor}.  The solid lines and
circles represent the input wires, the output wire and the `power'
wire.  The role of the power wire is to provide a growing tip for the
output if needed.  The dashed line represents a walk that will stick
at one of the three open circles labeled `a,' `b' or `c.'  The dashed
walk evaluates the gate and so must not occur until the input and
power wires have been grown to completion.  Suppose that input 1 is
{\sc true} so that the corresponding segment of wire is occupied.
Then the trajectory sticks at `a.'  If input 1 is {\sc false} but
input 2 is {\sc true}, the dashed walk sticks at `b.'  Finally, if both
inputs are {\sc false} then the walk sticks at `c.'  The occupancy of
site `c' records the output of the gate.

Fig.~\ref{dlaor} shows the implementation of a single input {\sc or}
gate.  Effectively, it shows how to cross the power wire (running
toward the left) over a logical wire (running vertically).  First,
the logical wire is grown to site `1.'  The walk represented by the
dotted line on the right sticks at `a' if the logical wire is {\sc
true} and sticks at `b' if the logical wire is {\sc false}.  Similarly,
the walk represented by the dotted line on the left sticks at `c' if
the logical wire is {\sc true} and sticks at `d' if the logical wire
is {\sc false}.  Finally, the logical wire may continue to grow
vertically and the power wire may continue to grow to the left without
interfering with one another.  In this way a single input {\sc or}
is simulated.

For three and higher dimensional lattices each gate may be separately
supplied with its own power wire.  For example, each gate has a
`column' dedicated to its power wire.  When the wire reaches the
appropriate height, it is routed horizontally to the desired gate.
For two dimensional DLA there is an additional complication in
arranging to have the power wire arrive at each gate without
interfering with the wires that carry truth values.  To accomplish
this we use a single power wire for the entire circuit.  It is
`snaked' through the gates level by level.  See the power wire in the
example shown in Fig.~\ref{dlalevel}; this example is dicussed further
later.

The discussion above shows how to pass the power wire through an {\sc
or} gate.  It is also necessary to have the power wire cross through a
{\sc nor} gate.  This can be done as shown in Fig.~\ref{dlacross}.  In
this figure the path of the power wire is numbered and the walks that
bring particles to the wire are shown as dashed lines.  Recall that
exactly one of the sites `a,' `b' or `c' is occupied during the
evaluation of the gate.  If `c' is occupied, the power wire grows
along the full path `1$-$8.'  If instead `b' is occupied, sites `1'
and `2' are skipped and growth starts at `3.'  In this case the walks
that go to `1' and `2' turn around there and return to the upper
boundary where they are absorbed.  Finally, if `a' is occupied then
growth of the power wire starts at `6.'  Thus after helping with the
evaluation of a {\sc nor} gate, the power wire may be passed through.

A single power wire traverses all the gates in the sequence in which
they would be evaluated in topological order.  This requires that the
gates be arranged in levels as shown in the example in
Fig.~\ref{dlalevel}.  The thick lines are circuit wires, the filled
circles are {\sc nor} gates and the open circles are single input
{\sc or} gates.  The power wire traverses the gates one level at a
time.  Gates are evaluated from bottom to top and level by level along
the path of the power wire.  The lower edge of the lattice is used as
a source for {\sc true} inputs to gates.  At each gate the power and
input wires arrive first, then the gate is evaluated and finally the
power wire is continued to the next gate.  After an entire level has
been simulated, outputs are grown to the succeeding level.  The
routing between levels can be accomplished by `spreading' the circuit
out on the lattice and then allocating a couple of horizontal channels
to each output of a gate.  An output will be grown upward to its
designated channel, grown horizontally underneath its appropriate gate
and then grown upward to serve as an input.  The planarity of the
original circuit guarantees that there will be no interference of
walks during this routing.
\comment{If there are n gates use 4n levels in between to do the
routing.  2/7/95 RG}

The reduction described above shows that the special instance of CVP
we constructed is faithfully evaluated by the growth of the DLA
cluster.  Furthermore, it is an {\bf NC} reduction.  The key point is
that the choice of paths for the walks is independent of the
evaluation of the circuit.  The full layout of the walks is given
globally by the planar layout of the original circuit as outlined
above and locally by Figs.~\ref{dlanor} through~\ref{dlacross}.  All
calculations required to compute these walks can be performed in {\bf
NC}.  \hfill $\Box$

\subsection{Metropolis dynamics for the Ising model}

Configurations of the Ising model are defined by spin variables,
$\sigma_i$, on a lattice where each spin may take the value $-1$ or
$+1$. The conventional way to obtain equilibrium states of the Ising
model is via the Metropolis Monte Carlo algorithm.  At each step of
the algorithm a site $i$ is chosen at random and the energy change,
$\Delta E_i$, for flipping the spin at this site is computed.  The
energy change is given by
\begin{equation}
\Delta E_i= 2J\sigma_i \sum_{<i,j>} \sigma_j
\end{equation}
where the summation is over nearest neighbors of site $i$ and $J$ is
the coupling energy.  If $\Delta E_i \leq 0$ the spin is `flipped'
($\sigma_i \rightarrow -\sigma_i$), whereas if $\Delta E_i > 0$ the
spin is flipped with probability $e^{-\Delta E_i/T}$, where $T$ is the
temperature.  After this procedure has been iterated sufficiently many
times, the resulting probability distribution for the spin
configurations is close to the equilibrium state.

Metropolis dynamics is governed by a random list of sites and, for
each site in the list, a random number $x_i$ with $0 \leq x_i<1$ such
that the site is flipped if $x_i \leq e^{-\Delta E_i/T}$.  We can
define the following natural decision problem for Metropolis
dynamics.\\

\noindent
{\bf Metropolis dynamics} (dimension $d$)\\
{\bf Given:} A positive integer $L$, an initial configuration of $L^d$
spins $\{\sigma_i\}$ with $\sigma_i \in \{ -1,+1\}$, a temperature
variable $Q=e^{-4J/T}$ where $Q$ is expressed as a $b$-bit binary
number, a designated site $s$ expressed in unary with $|s| = L^{d}$, a
list of $M$ sites and a list of $M$ random numbers $x_i$ with $0 \leq
x_i<1$ expressed as a $d b$-bit number.\\
{\bf Problem:} Is $\sigma_s= +1$ after running the Metropolis
algorithm?\\

\comment{
The reason for the $d$ times $b$ bit number above is so that
the comparisons given below can be done.

A small $Q$ value indicates large groups of $+$ or $-$ whereas
a value of say $Q = 1$ means the system is random. RG 2/14/95}

Given the random numbers $x_i$ we can assign {\em flip variables},
$g_i \in \{0,\ldots, d\}$, to each site $i$.  For example, in three
dimensions the flip variables are defined by the inequalities
\[
\begin{array}{llll}
g_i=0 & {\rm if} & 0 \leq x_i\leq Q^3, &\\
g_i=1 & {\rm if} & Q^3<x_i\leq Q^2, &\\
g_i=2 & {\rm if} & Q^2<x_i\leq Q & {\rm and}\\
g_i=3 & {\rm if} & Q<x_i < 1. &
\end{array}
\]
If a site $k$ is chosen for a possible flip at step $i$ and $\Delta
E_k /4J \leq 3-g_i$, then the flip is carried out; otherwise, the spin
is not changed. In other words, a chosen spin $i$ will flip at step
$j$ if it has $g_i$ or more neighbors of the opposite sign.  It is
clear that the Metropolis decision problem can be {\bf NC} reduced to
a version in which the random input is expressed as a list of flip
variables; it is this variant of the problem that we show is {\bf
P}-complete.

\begin{theorem}
\label{th-md}
Metropolis dynamics is {\bf P}-complete for $d$ greater than or equal
to 3.
\end{theorem}

{\em Proof sketch:\/} The Metropolis problem is proved {\bf
P}-complete by a reduction from monotone CVP\@.  The circuit is first
`embedded' in a three dimensional lattice.  The {\sc and} and {\sc or}
gates are represented by sites and wires connecting gates by chains of
sites and bonds.  For a circuit having $N$ edges, it can be shown that
such an embedding may be carried out in {\bf NC}\@.  Initially, all
spins on the lattice are $-1$. Logical values are represented by spin
values with $+1$ ($-1$) meaning {\sc true} ({\sc false}). Logical
values are propagated along wires by the following device: spins along
the wire are sequentially chosen for flipping and assigned the flip
variable 1. If the predecessor spin along the wire is $+1$, the
current spin will flip to $+1$ but if the predecessor spin along the
wire is $-1$, the flip is rejected.  Thus, once initiated, logical
values propagate along wires.  Wires must always be separated by one
or more lattice spacings except where they meet at gates.  Sites
representing gates have two input wires and two output wires.  After
all sites along the input wires have taken their logical values, the
gate is ready for evaluation. Gates are
assigned the flip variable 2 (1) for an {\sc and} ({\sc or}) gate.
Thus, if at least one input is {\sc true} an {\sc or} gate registers
{\sc true}, while both inputs must be {\sc true} for an {\sc and} gate
to register {\sc true}\@.  This {\bf NC} reduction shows that we can
simulate an arbitrary monotone circuit using Metropolis dynamics in
three or more dimensions.  Therefore, the Metropolis dynamics problem
is {\bf P}-complete.  \hfill $\Box$

Note that the planar monotone circuit value problem is in {\bf NC},
see~\cite{GrHoRu} for a list of references regarding this problem.
So, our proof does not show that the two-dimensional Metropolis
problem is {\bf P}-complete.  We have been unsuccessful in our
attempts to implement a {\sc not} gate within the framework of
Metropolis dynamics.

The construction in Theorem~\ref{th-md} relies on a special ordering
of the sites chosen for flipping.  However, we can easily extend the
proof to updating schemes in which sweeps through the lattice are
performed in a fixed order.  For example, consider the case of
parallel updating where first the odd sublattice is flipped all at
once and then the even sublattice.  The problem statement is slightly
different here since now at each time step flip variables are assigned
to half the sites in the lattice.  It is easy to keep sites inactive
by assigning them flip variables 3.  Sites are assigned flip variables
1 or 2 as in the above construction at the times they are to be
evaluated.

\subsection{Cluster dynamics for the Ising model}

Cluster flipping algorithms due to Wolff~\cite{Wolff} and Swendsen and
Wang~\cite{SwWa} are very efficient methods for generating equilibrium
states of the Ising model near criticality.  In this section we show
that natural decision problems associated with the Wolff and
Swendsen-Wang algorithms are {\bf P}-complete.

We illustrate the Wolff algorithm on an $L
\times L$ square lattice.  The starting point is a configuration of
spins, $\{\sigma_j\}$.  Next the bonds of the lattice are
independently occupied with probability $p$ as in bond percolation.
The occupation parameter is related to the temperature, $T$, according
to $p$~$=$~$1 - Q$ with $Q=e^{-2J/T}$ and $J$ the coupling energy
between neighboring spins.  A site $u$ on the lattice is chosen at
random and a cluster is grown from this site.  A site $v$ is in the
cluster grown from $u$ if there is a path from $u$ to $v$ such that
all the bonds along the path are occupied and all the spins along the
path including $\sigma_v$ are equal to $\sigma_u$.  The cluster of
spins defined in this way is `flipped' ($\sigma
\rightarrow -\sigma$ for each $\sigma$ in the cluster) which yields a
new spin configuration.  The procedure is iterated $M$ times. If the
temperature $T$ is chosen to be the critical temperature and if $M$
is sufficiently large the final configuration of spins is close to
the equilibrium Ising critical point. At the Ising critical
temperature, the clusters defined by the algorithm are critical
droplets~\cite{Fish67,FoKa,CoKl}
with Hausdorf dimension, $D_H$ equal to $15/8$.

The Swendsen-Wang algorithm is very similar to the Wolff algorithm
except that in each step of the algorithm {\em all\/} connected
clusters defined by the occupied bonds are identified.  All sites of
each cluster are assigned the same spin value.  The spin values for
each cluster are determined independently by a fair coin toss.

For each iteration of the Wolff or Swendsen-Wang algorithm, every bond
of the lattice is occupied with probability $p$ equal to $1-Q$\@.  To
implement this we utilize random numbers $x_{ij}$ with $0 \leq
x_{ij}<1$ for each nearest neighbor pair $(ij)$.  The bond $(ij)$ is
occupied if $x_{ij}$ is greater than $Q$\@.  At each time step a
cluster is grown from the starting point according to the occupation
variables and the current spin configuration as described above.  This
cluster is flipped and the procedure repeated $M$ times. We can define
the following natural problem based on Wolff dynamics.\\

\comment{Throughout any non-zero $Q$ will work.  Since $Q$ is
$e$ to some power we are always guaranteed it will be non-zero.
2/14/95 RG.

As $T$ goes to infinity, $p$ goes to 0 and we get isolated sites.
As $T$ goes to 0, $p$ goes to 1 and we get large connectivity.
At the critical point (a temperature), there is a fractal that spans
the lattice.  2/15/95 RG}

\noindent
{\bf Wolff dynamics} (dimension $d$)\\
{\bf Given:} A positive integer $L$,
an initial configuration of $L^d$ spins
$\{\sigma_i\}$ with $\sigma_i \in \{ -1,+1\}$,
a temperature variable $Q=e^{-2J/T}$ where $Q$ is
expressed as a $b$-bit binary number, a designated site $s$ expressed
in unary with $|s| = L^{d}$, a list of $M$ sites and $dML^d$
random numbers $x_{ij}$ with $0 \leq x_{ij}<1$ expressed as a $b$-bit
number.\\ {\bf Problem:} Is $\sigma_s = +1$ after running the
Wolff algorithm?\\

Given the random numbers $x_{ij}$ we can assign {\em bond occupation
variables\/} $b_{ij}$ such that $b_{ij}=0$ if $x_{ij} \leq Q$ and
$b_{ij}=1$ otherwise.  Bonds are counted as occupied if $b_{ij}=1$.
It is clear that the Wolff decision problem can be {\bf NC} reduced to
a version in which the random input is given as the $b_{ij}$ instead
of the $x_{ij}$.  It is this version that we show is {\bf P}-complete
using a reduction from the planar {\sc nor} circuit value problem.

\begin{theorem}
\label{th-wd}
Wolff dynamics is {\bf P}-complete.
\end{theorem}

{\em Proof idea:\/} The reduction is best illustrated by a simple
example.  We sketch it for the case $d$ equals two.  Consider the
planar circuit shown in Fig.~\ref{fignor} with three inputs and three
{\sc nor} gates.  The evaluation of this circuit can be reduced to the
Wolff problem shown in Fig.~\ref{figwolff}.  The lower case letters
indicate occupied bonds and time steps that are arranged in
alphabetical order.  All bonds labeled `a' are occupied only during
step 1, all bonds labeled `b' are occupied only during step 2 and so
on.  Bonds that are not explicitly shown are never occupied.  All
spins on the lattice originally have the value $+1$ except those that
are labeled {\sc false}\@.  A $+1$ spin represents {\sc true} and
vice versa. Numbers label initiation sites for cluster growth.
Cluster growth is initiated at gates and logical constants.  Site `1'
initiates the cluster growth at time step 1 and represents the {\sc
true} input to the circuit; site `2' initiates growth at time step
`2' and so on.  The first cluster propagates as far as site `4' and
both sites `1' and `4' (and the intermediate site) are flipped to
$-1$.  Site `2' initiates the next cluster which does not propagate.
Site `2' is flipped to $+1$. The cluster initiated at `3' does not
propagate so that at the beginning of time step 4 site `4' is in the
{\sc false} state; the first gate has been properly evaluated.  At time
step 4 site `4' flips but nothing else happens; the second
gate properly evaluates to {\sc true}\@.  At time step 5 a
cluster propagates from site `5' to the output gate that is flipped
to $-1$.  The output of the circuit is {\sc false} as it should be.

More generally, a {\sc nor} gate is represented by spins, and wires
connecting gates are represented by paths of bonds and spins.  All of
the bonds in a wire are occupied at the time step during which the
wire transmits its logical value.  If the logical value is {\sc true},
a cluster of up spins is propagated along the wire and the output end
of the wire is flipped to {\sc false} if this has not yet occurred.  A
gate transmits its value by initiation of a cluster and the output of
a gate can be read off as soon as all of its predecessors have
transmitted their values.  Note that a fan-out higher than two
may easily be supported.  Sites representing gates and logical
constants must not be nearest neighbors.  It
is clear that the reduction of the circuit to Wolff dynamics can be
carried out locally and is an {\bf NC} reduction.  Since planar {\sc
nor} CVP is {\bf P}-complete, so is the Wolff dynamics problem.
\hfill $\Box$

\comment{Use the `diagonal' version of planar NOR CVP.  See
description in GrHoRu under PCVP.  2/14/95}

Next we turn our attention to a natural decision problem associated
with the Swendsen-Wang algorithm.  The problem statement requires
random `bits,' $c_i$ equal to $\pm 1$, to be used to determine the
spins in the clusters.  Sites are given a conventional
ordering. Connected clusters defined by $Q$ and the variables $x_{ij}$
are labeled by the lowest ordered site, $l$, in the cluster and all
the spins in the cluster are assigned the value $c_l$.\\

\noindent
{\bf Swendsen-Wang dynamics} (dimension $d$)\\
{\bf Given:} A positive integer $L$, an initial configuration of $L^d$ spins
$\{\sigma_i\}$ with $\sigma_i \in \{ -1,+1\}$,
a temperature variable $Q=e^{-2J/T}$ where $Q$ is
expressed as a $b$-bit binary number, a designated site $s$ expressed
in unary with $|s| = L^{d}$, a number of iterations $M$, a list of
$dML^d$ random numbers $x_{ij}$ with $0 \leq x_{ij}<1$ expressed as a
$b$-bit number and a list of $ML^d$ random bits, $c_i$.\\
{\bf Problem:} Is $\sigma_s = +1$ after running the Swendsen-Wang
algorithm?

\begin{theorem}
\label{th-sw}
Swendsen-Wang dynamics is {\bf P}-complete.
\end{theorem}

{\em Proof hint:\/} The proof is similar to that for the Wolff problem
and consists of a reduction from planar {\sc nor} CVP\@.  Here again a
value of $d$ equal to two suffices for the reduction.  Consider a
two-dimensional lattice and suppose that the conventional ordering of
sites on the lattice is from left to right and then from bottom to
top.  The occupation variables $b_{ij}$ are chosen the same as for the
reduction to the Wolff problem.  The cluster spin variables, $c_i$,
are $-1$ for gates and {\sc true} inputs at the time they transmit
their values and $+1$ for all other sites and all other times. \hfill
$\Box$

For both Wolff and Swendsen-Wang dynamics, a single iteration of the
algorithm can be accomplished in poly-logarithmic parallel time using
a polynomial number of processors.  This is because the most complex
step is the identification of a connected component(s), which can be
carried out by a standard {\bf NC} algorithm~\cite{GiRy}.  More
specifically, if the number of iterations $M$ is set to a constant in
the statement of either the Wolff problem or the Swendsen-Wang
problem, the resulting decision problem is in {\bf NC}\@.  For the
Metropolis algorithm, an even stronger result holds.  If $M$ equals a
constant, the Metropolis decision problem is in {\bf AC}$^0$.  These
conclusions are not in conflict with the {\bf P}-completeness proofs
that rely on setting $M$ comparable to the size of the Boolean circuit
being simulated.  The {\bf P}-completeness results show that a
polynomial (in the system size) number of iterations of these
algorithms cannot be compressed into poly-logarithmic
number of parallel steps, unless {\bf NC}~$=$~{\bf P}.

\section{Discussion}
\label{se-discussion}

\subsection{Summary of results}

We have studied the computational complexity of natural decision
problems associated with several models in statistical physics.  Our
results can be summarized as follows:

\begin{enumerate}

\item
Mandelbrot percolation is in (non-uniform) {\bf AC}$^0$
(Theorem~\ref{th-mp}).

\item
Diffusion limited aggregation is {\bf P}-complete (Theorem~\ref{th-dla}).

\item
Metropolis dynamics for the Ising model is {\bf P}-complete
(Theorem~\ref{th-md}).

\item
Wolff dynamics for the Ising model is {\bf P}-complete
(Theorem~\ref{th-wd}).

\item
Swendsen-Wang dynamics for the Ising model is {\bf P}-complete
(Theorem~\ref{th-sw}).

\end{enumerate}

\subsection{Scope of the results}

It is important to understand the limitations of the {\bf
P}-completeness results for DLA and the variants of Ising dynamics.
Suppose we accept the hypothesis from complexity theory that {\bf
NC}~$\neq$~{\bf P}\@.  In this case the particular dynamics discussed
here for generating DLA clusters or equilibrium Ising configurations
are inherently sequential and cannot be efficiently simulated in
parallel. There are other ways to generate (approximate) equilibrium
states of the Ising model or DLA clusters; our results do not imply
that these ways are associated with {\bf P}-complete problems.
However, it was previously shown~\cite{Mac93a} that a second method of
producing DLA clusters is also {\bf P}-complete.  It seems plausible
that there are no poly-logarithmic time methods for sampling the DLA
distribution.  On the other hand, the jury remains out on whether it
is possible to sample from an approximation to the equilibrium
critical distribution for spin models in poly-logarithmic
time.  It would be of great interest to obtain results on the
difficulty of sampling physically interesting distributions.

A second limitation of the {\bf P}-completeness statements is that
they are worst case rather than average case results.  For example,
assuming that {\bf NC}~$\neq$~{\bf P}, we know that there exist
instances of the DLA problem that cannot be solved in poly-logarithmic
time although we do not know whether these `hard' instances are
typical or very rare.  Indeed, the instances used in the {\bf
P}-completeness proofs are atypical.  If the `hard' instances are
sufficiently rare, we may be able to sample the distribution in
poly-logarithmic time on average.  The theory of average case
complexity~\cite{Le86,BeChGoLu} addresses questions of this kind.
Unfortunately, it is not easy to see how to apply this theory to the
present problems.

\subsection{Parallel complexity and critical slowing down}

Away from a critical point, the equilibration time of real systems
without macroscopic inhomogeneities is independent of system size.
Similarly, the Metropolis algorithm can generate good approximations
to equilibrium configurations of the Ising model away from the
critical point in constant parallel time since each sweep can be done
in constant time and the number of sweeps is independent of the system
size. The associated decision problem is in {\bf AC}$^0$.

Equilibration of many real systems becomes increasingly slow near
critical points.  Typically the equilibration time at a critical point
scales as $L^z$, where $L$ is the system size and $z$ is the dynamic
exponent.  This phenomena, known as critical slowing down, also
afflicts most Monte Carlo methods used to sample spin configurations
at critical points.  The dynamic exponent $z$ is customarily defined
for Monte Carlo dynamics if relaxation to equilibrium requires
flipping $L^{z+d}$ spins.  It is often said that an algorithm suffers
no critical slowing down if $z = 0$ (with possible logarithmic
corrections).  This is not a satisfactory general definition of
`absence of critical slowing down.'  For example, imagine an algorithm
for which each spin is flipped only once but an enormous computation
is required to decide whether or not to effect the flip.
Alternatively, one might propose that `absence of critical slowing
down' means that the sequential time (computational work) is
$o(L^{d+\epsilon})$ for any $\epsilon > 0$.  This definition is both
machine dependent and unnecessarily stringent.

We propose that `absence of critical slowing down' should be
identified with the class {\bf NC}\@.  A sampling method suffers no
critical slowing down if it can be run in poly-logarithmic time on a
P-RAM with polynomially many processors.  This definition is, for the
most part, in agreement with the $z=0$ definition.  If $z>0$ for the
Monte Carlo methods studied here, the {\bf P}-completeness results
show that there is critical slowing down according to the new
definition.  On the other hand, if $z=0$ for either the Metropolis or
Swendsen-Wang algorithms, there is no critical slowing down since a
single sweep through the lattice for either of these algorithms can be
done efficiently in parallel and $z=0$ implies a poly-logarithmic
number of sweeps.  In contrast, the Wolff algorithm suffers critical
slowing down even for $z=0$ according to the new definition.  The
reason is that the average size of Wolff clusters scales as
$L^{\gamma/\nu}$, where $\gamma$ is the susceptibility exponent and
$\nu$ the correlation length exponent.  Thus, even if $z=0$ one
typically requires $L^{d-\gamma/\nu}$ iterations of the algorithm to
reach equilibrium. The {\bf P}-completeness result shows that carrying
out these iterations can almost certainly not be done in
poly-logarithmic time using a polynomial number of processors.

\subsection{Final remarks}

In this and two previous papers~\cite{Mac93a,MaGr} we have
investigated the parallel computational complexity of a variety of
models in statistical physics.  We have claimed that parallel
complexity provides statistical physics with a robust and sharply
defined measure that reflects some of our more intuitive notions of
complexity.  We have classified a wide variety of models into three
broad classes: those that require constant parallel time to simulate,
those that require poly-logarithmic time and those that require
polynomial time.  In each case we allow a polynomial number of
processors.  Even among models that generate random fractal patterns,
we find representatives in each of these classes.  Comparisons between
members of different classes reveal that models in the higher classes
generally pose a more difficult theoretical challenge.  It would be
extremely interesting to find more precise correlations between
computational complexity and the quantities conventionally studied in
statistical physics.

\section*{Acknowledgements}

Jon Machta is supported in part by National Science Foundation
Grant~DMR-9311580 and Ray Greenlaw by National Science Foundation
Grant~CCR-9209184.

\newpage
\section*{List of Figures}
\begin{enumerate}
\item
A realization
of Mandelbrot percolation.
\label{manperc}
\item
A realization of diffusion limited
aggregation.
\label{dlareal}
\item
Gadget for a {\psc nor} gate. Filled circles and
connecting bold lines show the two input wires, the output wire and
the power wire. The dotted line shows the path of the walk that
evaluates the gate. This walk terminates on site `a,' `b' or
`c.'
\label{dlanor}
\item
Gadget for a single input {\psc or} gate (effectively crossing a
power wire and a logical wire). The logical wire first arrives at
`1,' then the two dotted walks carry the power wire across the
junction, sticking at `a' and `c' respectively if the logical wire
is {\psc true} or at `b' and `d' respectively if the logical wire is
{\psc false}.
\label{dlaor}
\item
Passing the power wire through a {\psc nor} gate
after its evaluation. The growth of the wire follows the numbered
sites where the first occupied site is either `1,' `3' or `6'
depending on whether `c,' `b' or `a' is occupied.
\label{dlacross}
\item
Layout of {\psc
nor} (solid circles) and single input {\psc or} gates (open circles)
in levels with the power wire (thin line) traversing the gates in
the order in which they are evaluated.
\label{dlalevel}
\item
A simple planar {\psc nor}
circuit with three inputs and three gates used to illustrate
Theorem \ref{th-wd}.
\label{fignor}
\item
The Wolff problem that
simulates the {\psc nor} circuit. Numbered sites represent gates and
logical constants. Bonds with the same letter are occupied during
the same time step, bonds `a' during step 1, bonds `b' during step 2
and so on.
\label{figwolff}
\end{enumerate}

\end{document}